\documentclass[aps,pre,twocolumn,showpacs,superscriptaddress]{revtex4}
\usepackage{graphicx,color,epsfig}
\usepackage{amssymb,amsmath,amsfonts,wasysym}
\begin{document}

\title{Emergence and persistence of communities in coevolutionary networks}
\author{J. C. Gonz\'alez-Avella} 
\affiliation{Instituto de F\'isica, Universidad Federal do Rio Grande do Sul - 910501-970, Porto Alegre, Brazil} 
\author{M. G. Cosenza}
\affiliation{Grupo de Caos y Sistemas Complejos, Centro de F\'isica Fundamental, Universidad de Los Andes, M\'erida, Venezuela}
\author{J. L. Herrera}
\affiliation{Departamento de C\'alculo, Escuela B\'asica de Ingenier\'ia, Universidad de Los Andes, M\'erida, Venezuela}
\author{K. Tucci}
\affiliation{Grupo de Caos y Sistemas Complejos, Centro de F\'isica Fundamental, Universidad de Los Andes, M\'erida, Venezuela}

\begin{abstract}
We investigate the emergence and persistence of communities through a recently proposed mechanism  
of adaptive rewiring in coevolutionary networks.  
We characterize the topological structures arising in a coevolutionary network  subject to an adaptive rewiring process and 
a node dynamics given by a simple voterlike rule. 
We find that, for some values of the parameters describing the
adaptive rewiring process, 
a community structure emerges on a 
connected network. 
We show that the emergence of communities is associated to a decrease in the 
number of active links in the system, i.e. links that connect two nodes in different states. 
The lifetime of the community structure state scales exponentially with the size of the system.
Additionally, we find that a small noise in the node dynamics can sustain a diversity of states and a community structure in time in 
a finite size system. Thus, large system size and/or local noise can explain the persistence of communities and diversity in many real systems.
\end{abstract}
\pacs{89.75.Fb; 87.23.Ge; 05.50.+q}

\maketitle

\section{Introduction}
Many social, biological, and technological systems possess a characteristic 
network structure consisting of 
communities or modules, which are groups of nodes distinguished by having 
a high density of links between nodes of the same group and a comparatively 
low density of links between nodes of different groups \cite{Girvan,Fortunato,Porter,Kivel}. Such a network structure
is expected to play an important functional role in many systems. In a social network, communities might indicate factions, interest
groups, or social divisions \cite{Girvan}; in biological networks, they encompass entities having
the same biological function \cite{Spirin,Huberman,Guimera}; in the  World Wide Web they may correspond to groups of pages
dealing with the same or related topics \cite{Geraci}; in food webs they may identify compartments \cite{Stouffer}; 
and a community in a metabolic or genetic network might be related to a specific functional task \cite{Thiele}. 

Since community structure constitutes a fundamental feature of many networks, the development of 
methods and techniques for the detection of communities represents one of the most active research areas in network science 
\cite{Fortunato,Newman1,Danon,Fortunato2,Blondel,Good,Malla,Bassett}. 
In comparison, much less work has been done to address a fundamental question: how do communities arise in networks? \cite{Palla}.
 
Clearly, the emergence of characteristic topological structures, including communities, 
from a random or featureless network requires 
some dynamical process that modifies the properties of the links representing the interactions between nodes. 
We refer to such link dynamics as a \textit{rewiring process}. Links can vary their strength, or they can appear and disappear
as a consequence of a rewiring process. In our view, two classes of 
rewiring processes leading to
the formation of structures  
in networks can be distinguished: 
(i) rewirings based on local connectivity properties 
regardless of the values of the state variables of the nodes,  
which we denote as \textit{topological rewirings}; 
and (ii) rewirings that depend on 
the state variables of the nodes, where the link dynamics is coupled to the node state dynamics 
and which we call \textit{adaptive rewirings}.

Topological rewiring processes 
have been employed to explain the origin of 
small-world  and scale-free networks \cite{Barabasi,Watts}. These rewirings can lead to the appearance of community structures 
in networks with weighted links \cite{Kaski1} 
or by preferential attachment driven by local clustering \cite{PRX}.
On the other hand, there is currently much interest in the study of networks 
that exhibit a coupling between topology and  states, since many systems observed in nature can be
described as dynamical networks of interacting nodes where the connections and the states of the nodes affect each other and 
evolve simultaneously \cite{Zimmerman,Maxi,Rohl,GrossR,GrossL,Zanette,EPL}. 
These systems have 
been denoted as coevolutionary dynamical systems or adaptive networks and,  
according to our classification  above, they are subject to adaptive rewiring processes. 
The collective behavior of coevolutionary systems is determined by the competition  of the time scales 
of the node dynamics and the rewiring process.
Most works that employ coevolutionary dynamics have focused on the characterization of the phenomenon of network fragmentation 
arising from this competition.
Although
community structures have been found in some coevolutionary systems \cite{Bocaletti1,Bocaletti2,Kaski2,Mandra}, 
investigating the mechanisms for 
the formation of perdurable communities remains an open problem.

In this paper we investigate the emergence and 
the persistence of communities in networks
induced by a process of adaptive rewiring. 
Our work is based on a recently proposed general framework for coevolutionary dynamics in networks \cite{EPL}. 
We characterize the topological structures forming in a coevolutionary network having a simple node dynamics. 
We unveil a region of parameters 
where the formation of a supertransient modular structure 
on the network occurs. 
We study the stability of the community 
configuration 
under small perturbations of 
the node dynamics, as well as for different initial conditions of the system.

\section{Emergence of communities through an adaptive rewiring process}
We recall that a rewiring process in a coevolutionary network can be described in terms of two basic actions that
can be independent of each other: 
disconnection and connection between nodes \cite{EPL}. 
These actions may correspond to discrete connection-disconnection events, or to continuous 
increase-decrease strength of the links, as in weighted networks. 

Both actions in an adaptive rewiring process are, in general, based on some mechanisms of
comparison of the states of the nodes. 
The disconnection action can be characterized by a parameter $d\in [0,1]$, that measures 
the probability that two nodes in identical states become disconnected, and such that $1-d$ is the probability that two nodes 
in different states disconnect from each other.
On the other hand, the connection action can be characterized by another parameter $r\in [0,1]$ that describes the probability 
that two nodes in identical states become connected, and such that $1-r$ is the probability that two nodes in different states connect
to each other \cite{EPL}. 
In a social context, these actions allow the description of diverse  manifestations of 
phenomena such as  inclusion-exclusion,
homophily-heterophily, 
and tolerance-intolerance. 

To  investigate the formation of 
topological structures through 
an adaptive rewiring process,
we consider a  
random network of $N$ nodes having average degree  $\bar{k}$. 
Let $\nu_i$ be the set of neighbors of node $i$, possessing $k_i$ elements. 
The state variable of node $i$ is denoted by $g_i$. For simplicity, 
we assume that the node state variable is discrete, that is, $g_i$  can take any of $G$ possible options.
The states $g_i$ are initially assigned at random with a uniform distribution.
Therefore there are, on the average, $N/G$ nodes in each state in the initial random network.
We assume that the network is subject to a rewiring process whose actions are 
characterized by parameters $d$ and $r$.

For the node dynamics, we employ an imitation rule such as a voterlike model that has been used in several 
contexts \cite{Holme,Holley,Castellano,Kra}. This model provides a simple dynamics for the node state change
without introducing any additional parameter. 
Parameters of the node dynamics can modify the time scale of the change of state of the nodes \cite{Vazquez}; 
however, those parameters  should not produce qualitative changes in the global behavior of the system.

Then, the coevolution dynamics in this system is given by iterating these three steps: 
(1) Chose at random a node $i$ such that $k_i>0$. 
(2) Apply the rewiring process:  select at random 
a neighbor $j \in \nu_i$ and a node $l \notin \nu_i$. 
If the edge $(i,j)$ can be disconnected according 
to the rule of the disconnection action 
and the nodes $i$ and $l$ can be connected according to the rule of the connection action, 
break the edge $(i, j)$ and create the edge $(i, l)$.
(3) Apply the node dynamics: chose randomly a node $m \in \nu_i$ such that $g_i\neq g_m$ and set $g_i=g_m$.
This rewiring conserves the total number of links in the network.  
We  have verified that the collective behavior of this system is statistically invariant if steps (2) and (3) are reversed.

The parameters $N$, $\bar{k}$, and $G$ remain
constant. We also maintain fixed the ratio $\gamma\equiv N/G=10$. 

To study the dynamical behavior of 
the network topology, 
we consider the time evolution of several 
statistical quantities in the system for different values of the parameters $d$ and $r$. 
We characterize the integrity of the network by calculating the normalized (divided by $N$)
average size of the largest component or connected subgraph in the system, regardless of the states of the nodes, at time $t$
denoted by $S(t)$, where a time step consists of 
$N$ iterations of the algorithm. 
We call a domain a subset of connected nodes that share the same state, 
and denote by
$S_g(t)$ the normalized average size of the largest domain in the system
at time $t$.  
Additionally, 
we calculate the fraction of links that are active in the system at a given time, that we call $\rho(t)$. 
A link is active if it connects two nodes in different states. Lastly, as a measure of the modular structure of the network, we 
define the quantity $\Delta Q(t) \equiv Q(t)-Q(0)$ as the modularity change, where $Q(t)$ is the modularity of the network at time $t$,
calculated through a community detection algorithm \cite{Blondel}, and $Q(0)$ is the value of this quantity for the initial random network.  

\begin{figure}[h]
\begin{center}
\includegraphics[width=0.76\linewidth,angle=0]{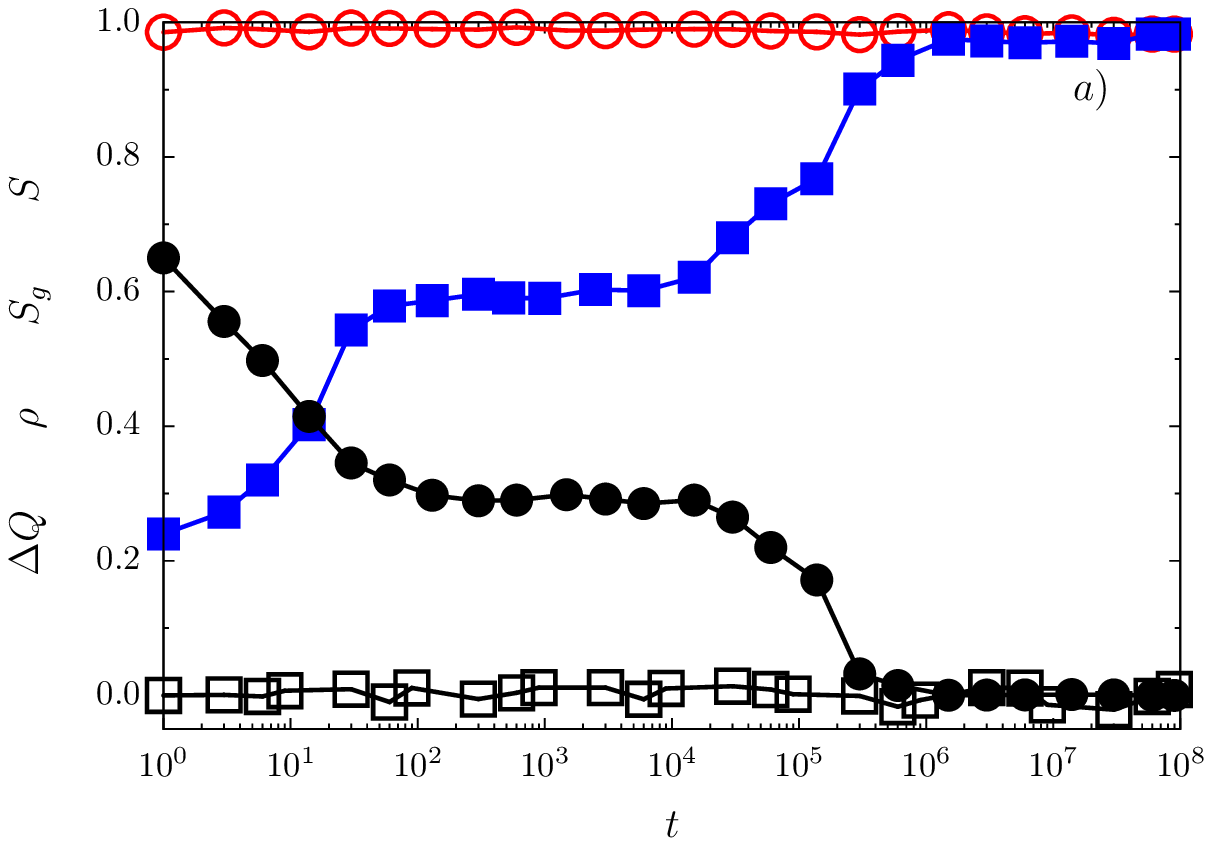}
\end{center}
\begin{center}
\includegraphics[width=0.76\linewidth,angle=0]{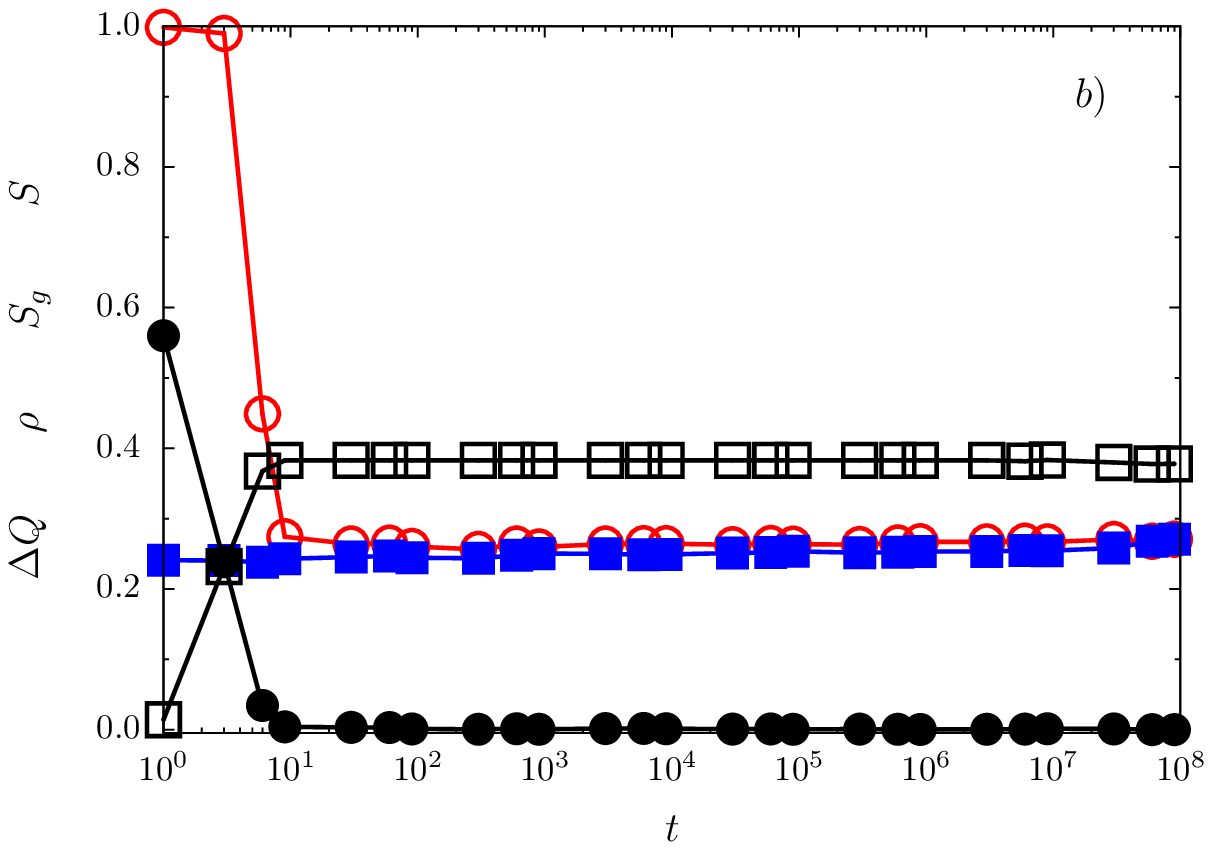}
\end{center}
\begin{center}
\includegraphics[width=0.76\linewidth,angle=0]{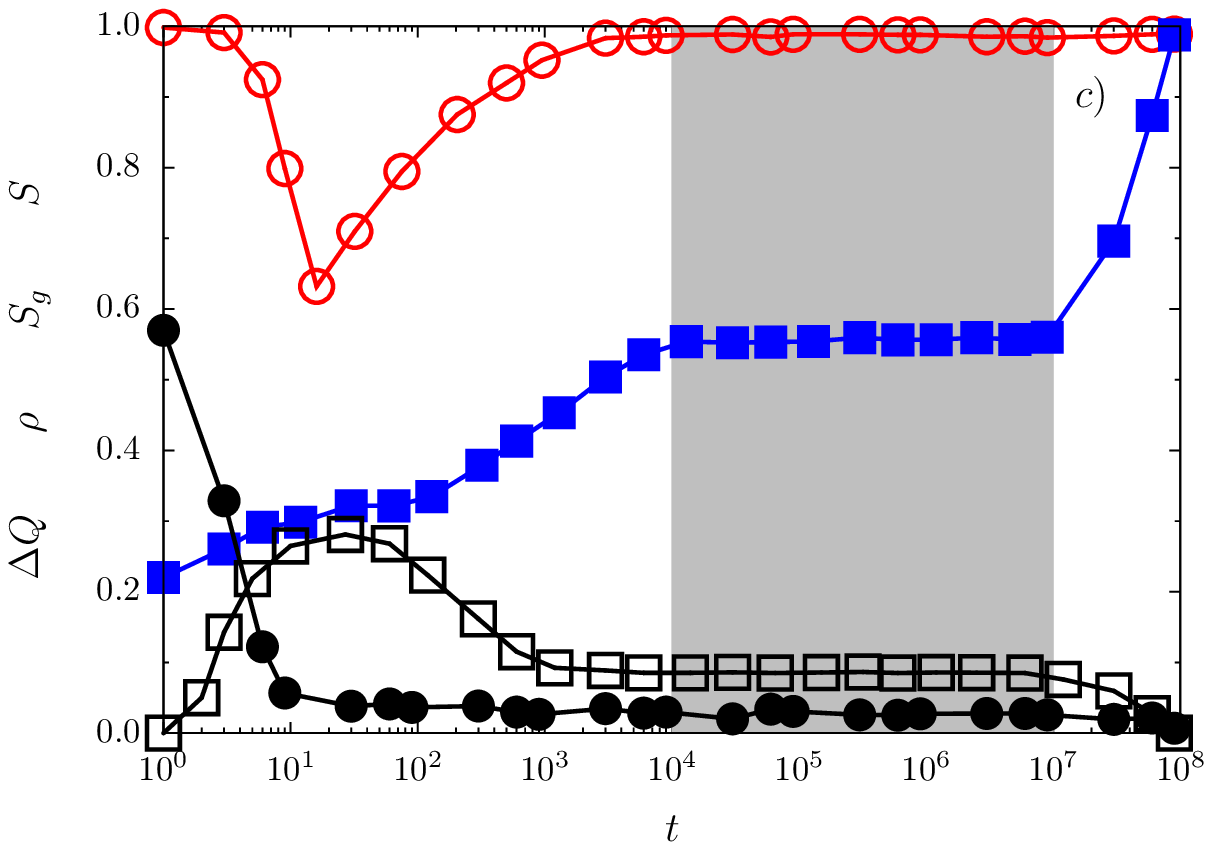}
\end{center}
\caption{Time evolution of the quantities $S$ ($\Circle$), $\rho$ ($\CIRCLE$), $S_g$ ($\blacksquare$)
and $\Delta Q$ ($\square$).  
System size is $N=80$, $\bar k=4$, and $G=8$; 
fixed parameter $d=0.2$. a) $r=0.2$. b) $r=1.0$. c) $r=0.8$.
The gray zone indicates the interval of time for which the quantity $\Delta Q$ reaches a constant value. 
All numerical data points are averaged over $20$ realizations of initial conditions.}
\label{F2}
\end{figure}

Figure~\ref{F2} shows the above four quantities as functions of time for a fixed value $d=0.2$ and different values of $r$.
For $r=0.2$, Fig.~\ref{F2}(a)  reveals 
that $S\rightarrow 1$ for all times, a value corresponding to
a large component whose size is comparable to that of the system. This indicates that the network remains connected 
during the evolution of the system. 
The quantity $S_g(t)$ initially increases in time until it reaches a stationary value $S_g(t) \approx 0.58$ during a long time interval 
(four orders of magnitude); there are two
connected groups of nodes in different states on the average. Due to finite size fluctuations \cite{Fede,ThiloN},  
the system eventually reaches a homogeneous absorbing state, where $S_g(t)=S \rightarrow 1$.
However, the sizes of these fluctuations decrease as the size of the system increases, until they decay to zero in the 
limit $N \rightarrow \infty$; 
in that situation the homogeneous absorbing state is not reached \cite{Fede}.
On the other hand, the fraction of active links $\rho(t)$ decreases as $S_g(t)$ increases, 
until $\rho(t)$ reaches a stationary value during the same interval of time as $S_g(t)$ becomes stationary. 
Since eventually one state survives 
on a large connected network component, the number of active links goes to zero. 
This behavior agrees with that observed in Refs. \cite{Fede,Maxi}. 
The value of the quantity $\Delta Q(t)$ remains close to zero, indicating that the modularity of 
the initial random network does not vary in time  
in this region of parameters. 

Figure~\ref{F2}(b) shows that, for $r=1$,  $S$ decays rapidly to a value tending to $\gamma/N$, indicating that the network has been fragmented 
in various small components. This fragmentation is associated 
with a rapid decay to zero of the fraction of active links $\rho$. The rapid drop of $\rho$ 
brings a limitation to the process of state change of the nodes and,
therefore, the size of the largest domain $S_g$ remains about the value of the average fraction
of nodes in a given state that are present in the initial network, i.e. $\gamma/N$. 
The fragmentation of the network is also reflected in the behavior
of $\Delta Q(t)$, that grows until a stationary value of maximum modularity
associated to the presence of separate domains, according to the employed algorithm \cite{Blondel}.

\begin{figure}[h]
\begin{center}
\includegraphics[width=0.6\linewidth,angle=90]{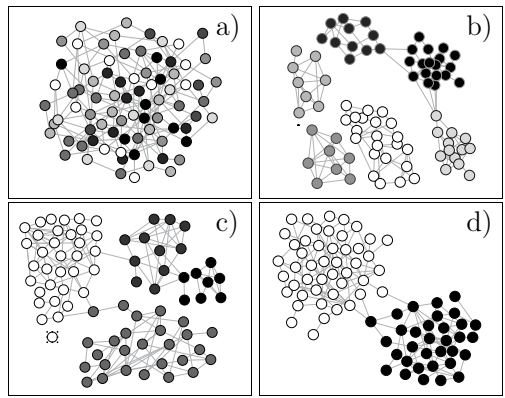}
\end{center}
\caption{Snapshots of the network structure and node states at different times during the evolution of the system, for one realization 
of initial conditions 
in the case of Fig.~1(c). Different node states are represented by different shades of gray. Fixed parameters are $N=80$, 
$\bar k=4$, $r=0.8$, $d=0.2$.
a) $t= 0$; b) $t=20$; c) $t=150$;  d) $t =10^5$.}
\label{F3}
\end{figure}

The evolution of the quantity $S(t)$ in Fig.~\ref{F2}(c) indicates
that the initial network with $S=1$ (visualized in Fig.~\ref{F3}(a)) undergoes a fragmentation process consisting of separated domains where $S$ 
decreases (Fig.~\ref{F3}(b)), 
and then a recombination process
takes place (Figs.~\ref{F3}(c), \ref{F3}(d)) until the network becomes a connected graph again, where $S \to 1$. A minimum value of $S$ separates 
these two
processes occurring during the time evolution of the system. The early fragmentation and recombination processes occurring in the network are also
manifested in the behavior of the  modularity change $\Delta Q(t)$, which exhibits a maximum as $S$ goes to a minimum. The minimum of $S$ also 
coincides 
with the decay of $\rho$ to a small value that is maintained for a long interval of time (four orders of magnitude in time, indicated in color gray), 
until eventually $\rho$ 
drops to zero when the nodes in the reconnected network reach a homogeneous state, corresponding to $S_g=1$. The subsistence of a minimum fraction of 
active links in the network for a long time permits the reattachment of 
separated domains to form a large connected network during this time interval, characterized by $S=1$ and
$S_g \approx 0.5$. Since
active links connect different domains, then the majority of links must lie inside the different domains coexisting on the large connected network. 
Therefore, there exist several domains inside which nodes are highly connected, with fewer connections between different domains.  
This type of network structure has been called a modular or community structure \cite{Girvan}. 
The corresponding network is visualized in Fig.~\ref{F3}(d). 
The emergence of a modular structure in the network is reflected in the quantity $\Delta Q(t)$, which
remains at a constant positive value during this stage. The asymptotic state of the system corresponds to a
large random connected network ($S=1$), similar to the initial one ($\Delta Q=0$), but with its nodes in a homogeneous state ($S_g=1$) and therefore, 
with no active links left ($\rho=0$).

\begin{figure}[h]
\begin{center}
\includegraphics[width=0.6\linewidth,angle=90]{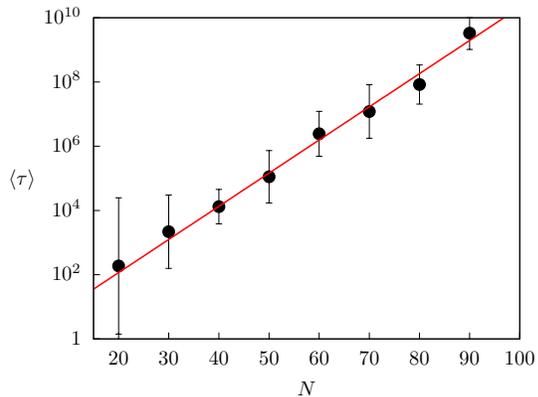}
\end{center}
\caption{Semilog plot of the average time $\langle\tau\rangle$ for which $\Delta Q(\tau)=0$, as a function of the system size $N$, 
for fixed values $\bar k=4$, $d=0.2$ and $r=0.8$.  The continuous line is the linear fitting with slope $\beta = 0.2 \pm 0.05$.
Error bars indicate standard deviations obtained 
over $10$ realizations of initial conditions for each point.}
\label{F5}
\end{figure}

To investigate the effects of the size of the system on the persistence of communities 
in the network, we show in Fig.~\ref{F5} 
a semilog plot of the average  asymptotic time  $\langle \tau \rangle$ for which $\Delta Q(\tau)=0$ ($\tau>0$), as a function of $N$.  
We numerically find that $\langle \tau \rangle$ scales exponentially with $N$ as $\langle \tau \rangle\sim e^{\beta N}$, 
with $\beta = 0.2 \pm 0.05$.  This behavior is characteristic of supertransient states in dynamical systems \cite{Kaneko}. 
For a finite size system, the 
modular structure and the coexistence of various domains on a connected network 
should eventually give place to one large domain. However, the asymptotic  random connected network in a homogeneous state cannot 
occur in an infinite size system. Thus, for large enough $N$, 
the decay of the modular structure cannot be observed in practice.

\begin{figure}[h]
\begin{center}
\includegraphics[width=0.7\linewidth,angle=90]{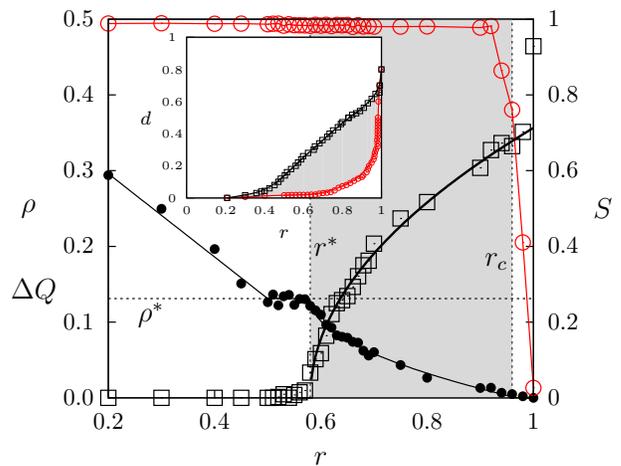}
\end{center}
\caption{$\Delta Q$ ($\square$), $\rho$ ($\CIRCLE$), and $S$ ($\Circle$, right vertical axis) 
as functions of $r$, with fixed $d=0.2$, at $t=10^6$ (within the interval of subsistence of communities) 
for a network with $N=1000$, $\bar k=4$.
The continuous thick line is the fitting of the values of $\Delta Q$ corresponding 
to the function $\Delta Q \propto (r-r^*)^{\nu}$, with $\nu \approx 0.50 \pm 0.01$. 
The horizontal dashed line marks the value $\rho^*$ below which modularity emerges.  
Gray color indicates the region of parameters 
where  communities appear in the connected network. 
All numerical data points are averaged over $10$ realizations 
of initial conditions. Inset: space of parameters $(d,r)$ showing in gray the region where communities appear
within the boundary curves $d(r^*)$ ($\square$) and $d(r_c)$ ($\Circle$).
}
\label{F4}
\end{figure}

The emergence of a modular structure 
can be characterized by calculating the value of the modularity change $\Delta Q$ 
at a fixed time 
(within the corresponding lapse of existence of communities) as a function of 
$r$ with a fixed
value of $d$, as shown in Fig.~\ref{F4}. There is a critical value  $r^*$ below which $\Delta Q$ is zero, reflecting the subsistence 
of the initial random 
topology, and above which  $\Delta Q$ increases,
indicating the appearance of a modular structure in the network.
The onset
of modularity 
can be described by the 
relation $\Delta Q \propto (r-r^*)^{\nu}$, with $\nu \approx 0.50 \pm 0.01$, 
typical of a continuous phase transition.  
Figure~\ref{F4} also shows the fraction of active links $\rho$ at $t=10^6$ as a function of $r$. We observe that 
the modularity transition at $r^*$ coincides with a drop of $\rho$ to small values below a value $\rho^*$. 
Since active links are associated to the contact points defining the interphase between different domains \cite{Mandra}, 
a low density of active links constrains the growth of domains,
giving rise to the modular structure in the network. 

In Fig.~\ref{F4} we also plot  
$S$ as a function of $r$. There is a critical value 
$r_c\approx 0.96$ above which a fragmentation of the network, 
characterized by  $S \rightarrow 0$, takes place. 
The employed modularity measure gives high values for $r >r_c$, manifesting the presence of  
trivial communities or separated graph components. 
We have verified that algorithm \cite{Newman1}
gives a
behavior for modularity similar to that shown in Fig.~\ref{F4} for $r \in [r^*,r_c]$. 
For $r<r^*$, we obtain $S \rightarrow 1$ and $\Delta Q =0$; indicating that the network remains 
connected and preserves its initial random structure. 
The modular structure appears in the connected network for 
$r^*<r<r_c$; 
this state is characterized by $S \rightarrow 1$, $\Delta Q >0$, 
and $\langle \tau \rangle\sim e^{\beta N}$.
The inset in Fig.~\ref{F4} shows the region on the space of parameters $(d,r)$ where communities appear.
Network fragmentation in this space
occurs for parameter values 
below the  open-circles boundary line.

\section{Stability of communities}
To shed light on the nature of the transient behavior of the modular structure, 
we introduce a perturbation in the node dynamics as follows: at each time step (every $N$ iterations of the algorithm) 
there is a probability $\xi$ that a randomly chosen agent changes its state assuming any of the $G$ possible states at random. 
Thus, the parameter 
$\xi$ represents the intensity of 
the random noise affecting the node dynamics, with $\xi=0$ corresponding to 
the original algorithm. Intrinsic random noise in the local states has been employed to simulate
the phenomenon of cultural drift in models of social dynamics \cite{Axelrod,Max}.
In addition, we study the robustness of the communities 
for different initial conditions of the system: 
(i) an initial random network and a random distribution of states; (ii) an initial random network and a homogeneous state; and 
(iii) an initial fragmented network consisting of $G$ separated domains, each with $N/G$ nodes. 
Condition (i) corresponds to the initial condition used in the original algorithm, while 
initial conditions (ii) and (iii) correspond to the absorbing states in the connected  and the fragmented configurations, 
respectively. 

\begin{figure}[h]
\begin{center}
\includegraphics[width=1.0\linewidth,angle=0]{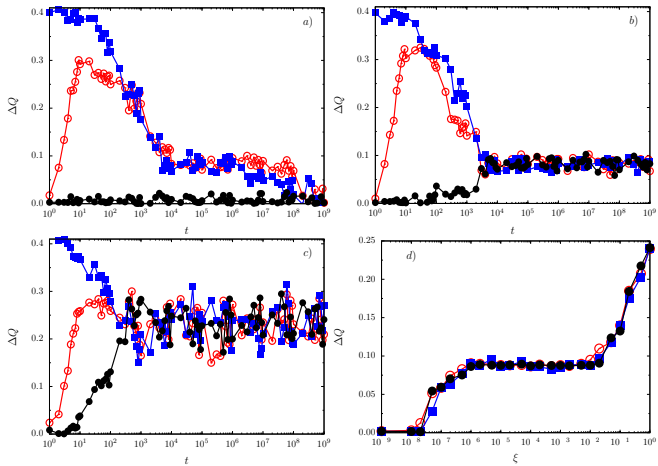}
\end{center}
 \caption{
Time evolution and dependence of $\Delta Q$ on the intensity of the noise $\xi$ for three different initial conditions of the
network structure and states, 
with fixed 
$N=80$, $\bar k=4$, $G=8$, $d=0.2$, $r=0.8$. On each panel,
$\Circle$ initial condition (i); $\CIRCLE$ initial condition (ii);  $\blacksquare$ initial condition (iii).
(a) $\Delta Q$ versus time with $\xi=0$. (b) $\Delta Q$ versus time with $\xi=10^{-4}$.  
(c) $\Delta Q$ versus time with $\xi=1$.  
(d) $\Delta Q$ as a function of $\xi$ at  $t=10^9$.}
\label{F7}
\end{figure}

Figures~\ref{F7}(a)-(c) show  $\Delta Q$ versus time with fixed parameters $d=0.2$, $r=0.8$,
for three different values of the intensity of the noise $\xi$ and the three initial conditions described above. 
Figure~\ref{F7}(a) shows that, in absence of noise and regardless of the initial conditions, the system reaches 
the same asymptotic state, with  $\Delta Q=0$, as in Fig. 1(c). No transient structures appear for the homogeneous initial condition (ii),
as expected; however 
a modular structure emerges as a transient state for conditions (i) y (iii). For these conditions, the transient time for the modular 
structure depends on the system size as in Fig.~3.  
Figure~\ref{F7}(b) shows that a modular structure, characterized by a nonvanishing value of $\Delta Q$, can be sustained in time by the
presence of a small noise for the different initial conditions, 
in spite of the finite size of the network.
We have verified that $S = 1$ for the three cases in both Fig.~\ref{F7}(a) and Fig.~\ref{F7}(b). 
A larger noise intensity leads to an increment of
the value of $\Delta Q$ for the different initial conditions, as shown in Fig.~\ref{F7}(c). For the three cases
we obtained $S<1$, corresponding to a fragmented network. 

Figure~\ref{F7}(d) shows $\Delta Q$ as a function of $\xi$ at fixed time $t=10^9$, after transients, for the three initial conditions.
Note that the asymptotic behavior of $\Delta Q(\xi)$ is independent of the initial conditions. There is
an intermediate range of the noise intensity where a modular structure can be maintained in the network. The value of 
$\Delta Q(\xi)$ in this region corresponds to the value of this quantity observed in the temporal plateau in Fig.~1(c). 

Our results show that, for an intermediate range of noise intensity, the modular structure 
can be sustained in time in a finite size coevolutionary system. 
An appropriate level of noise keeps the diversity of states in the system and prevents the disappearance of active links. 
As a consequence, the convergence to a homogeneous asymptotic state does not occur.
The role of noise in the modular configuration 
is similar to that of the limit of infinite system size, $N \to \infty$,
where a diversity of states is always present
and domains can subsist indefinitely.  

\section{Conclusions} 
We have employed a recent description of the process
of adaptive rewiring in terms of two actions: 
connection and  disconnection between nodes, both based on some criteria for comparison of the nodes state variables \cite{EPL}.
We have found that, 
for some values of the parameters $r$ and $d$ characterizing these actions, 
a  modular structure emerges previous to the settlement of a random network topology. 
The  actions  
of the rewiring process modify the competition between the time scales of the rewiring and the node dynamics, 
and therefore they can also control the emergence of communities.
The modular behavior separates 
two network configurations 
on the space of parameters $(d,r)$: 
a state where the initial random 
topology stays stationary in time, 
and a fragmented configuration. 
We have shown that the modular structure is a supertransient state. 

The presence of  communities has been characterized 
by several collective 
properties: the network is connected ($S \to 1$); there are various domains coexisting on the network ($S_g<1$); and the modularity measure
increases with respect to that of the initial random network ($\Delta Q >0$).

The formation of modular  structures is related to the number of active links present
in the network: communities emerge when the fraction of those links drops to small values. 
Since active links are associated with contact points that define the interphase between different domains in the network, a low density of 
active links means a restriction to the possibility of growth for domains. As a result, different domains are connected by few links, leading 
to the appearance of communities. 

The appearance of a short-lived modular structure always precedes the fragmentation of the network: 
Fig.~\ref{F2}(b) 
shows that the quantities $\rho$, $S$, $S_g$, and $\Delta Q$ at time $t=5$ reach those values associated to a modular structure. 
We have verified, by plotting successive snapshots, that the network topology indeed passes through a modular phase before becoming fragmented.
Thus, communities constitute temporary configurations that are likely to emerge during the evolution of 
the network topology of coevolutionary systems. 
Community structure has also been observed in 
the transient dynamics of models of epidemic spreading on adaptive networks \cite{Yang}.
We have found that,
for appropriate parameter values of the corresponding adaptive rewiring process, 
the community structure can become a supertransient state.

We have shown that noise in the node dynamics can sustain a diversity of states and the community structure in time 
in a finite size coevolutionary system. 
The role of noise on the lifetime of the modular structure state is similar to that of the limit of infinite system size. 
Thus, large system size and/or local noise can explain the persistence of communities and diversity in many real systems \cite{Libro,Area}.

\section*{Acknowledgements} 
J.C.G-A acknowledges support from CNPq, Brazil. M. G. C. is grateful to the Senior Associates Program of 
the Abdus Salam International Centre for Theoretical Physics, Trieste, Italy, for the visiting opportunities.

\end{document}